# Zero-field spin noise spectrum of an alkali vapor with strong spin-exchange coupling

Revised 6/11/2021  14:27:00

Ya Wen[1,2], Xiangyu Li[1,2], Guiying Zhang[3], Kaifeng Zhao[1,2]*

[1] Key Laboratory of Nuclear Physics and Ion-Beam Application (MOE), Fudan University, Shanghai 200433, China
[2] Institute of Modern Physics, Department of Nuclear Science and Technology, Fudan University, China
[3] College of Science, Zhejiang University of Technology, Hangzhou 310023, China
*email address: zhaokf@fudan.edu.cn

We study the zero-field optical spin noise (OSN) spectra for a thermal state dense $^{87}$Rb vapor. Our main findings are: (1) The OSN spectrum consists of two components representing a positive and a negative hyperfine spin correlation (HSC), the relative power of which varies dramatically with the detuning frequency of the probe. (2) There exist two polar frequencies at which the OSN spectrum is completely polarized with one type of HSC. (3) At the limit of far detuning, the power ratio of the positive and negative HSC component of the OSN is 4:5. (4) The total power of the OSN is independent of the strength of SE coupling. (5) We give a simple way of deriving the OSN using the eigensolution of the density matrix equation.

As a basic two-body interaction, spin-exchange (SE) between the valence electrons of two alkali atoms [1–4] or between the electron of an alkali atom and the nucleus of a noble gas atom [5–8] plays an important role in the physics of atomic clocks [9], atomic magnetometers [10], magnetic resonance imaging [11,12], precision measurements [13,14], and neutron physics [8].

During a SE collision, although the spin of an individual atom may flip randomly, the total spin of the colliding pair is conserved. In the presence of a magnetic field, SE causes relaxations of the collective spin component transverse to the field if the participants have different g-factors. However, in the strong SE coupling regime, where the SE rate far exceeds the Larmor frequency difference of the participants and other relaxation rates, all spins are coupled together and precess coherently as one unit without any SE relaxation [4,15]. This effect lies at the heart of many applications of quantum metrology, such as the spin-exchange relaxation free (SERF) magnetometer [16,17], rotation sensors [18,19], searching new forces [20–22], and spin entanglement and squeezing in ultracold alkali gases [23–27].

While ultracold alkali atoms will remain on the lower hyperfine multiplet of the ground state under SE [23], the alkali atoms in vapor cells jump randomly between the hyperfine multiplets after a SE collision because their thermal energy greatly exceeds the hyperfine splitting. Thus, the earlier successes of spin squeezing and entanglement in vapor cells are achieved in low-density vapors with negligible SE [28–30]. In the pursuit of larger numbers of squeezed atoms and longer entanglement times, the strong SE coupling regime has attracted more attention recently. Studies include spin entanglements created or preserved by SE [31–33], the proposal of SE mediated nuclear spin entanglement [34–37], and SE induced spin noise (SN) correlations between atoms of different alkali species [38–40]. However, despite the intense research, the SN of a single alkali isotope in the strong SE coupling regime has never been precisely studied. The SE effect on the hyperfine spin correlation (HSC) has been either ignored or oversimplified as a spin temperature distribution.

Optical spin noise (OSN) spectroscopy measures the SN-induced Faraday rotation (FR) fluctuations of an off-resonant probe [41–44]. Besides offering a non-perturbative detection of the dynamics of a spin system, OSN of the thermal state provides a robust tool to calibrate the standard quantum limit and the degree of spin squeezing [45]. The OSN is not always proportional to the SN in multiple optical resonances, and the dependence of the OSN power on the probe's detuning is sensitive to spin correlations [46].

This work studies the OSN spectra of a thermal state dense $^{87}$Rb vapor in the strong SE coupling regime. We measure the noise spectra in a π-pulse-modulated (πPM) transverse magnetic field [47,48], which shifts the SN resonance out of the swamp of 1/f noises while acting like a zero-field for the SE interactions [48]. We eliminate the diffusion distortion of the spectra by expanding the probe beam over the cell's entire cross-section [49] and using an evacuated antirelaxation coated cell. As the wall relaxation rate in such cells is much lower than the SE rate, we can observe the pure effect of SE on the OSN spectra. Also, without buffer gas, the optical-transition linewidth is much smaller than the ground-state hyperfine splitting. Thus, we can access smaller detunings and discover the 'polar' frequencies at which opposite kinds of HSCs are markedly revealed.

We first give a simple theory of the OSN for our system. Assuming the magnetic field points in the $x$ direction and SE is the only spin interaction, the master equation of the system's density matrix is

$$\frac{d\rho}{dt} = \frac{W[\mathbf{I}\cdot\mathbf{S},\rho]}{(I+1/2)i} + \frac{\omega_e[S_x,\rho]}{i} + \Gamma\left[\varphi(1+4\langle\mathbf{S}\rangle\cdot\mathbf{S})-\rho\right] \quad (1)$$

where $W$ is the ground-state hyperfine splitting, $\mathbf{S}$ the electronic spin, $\mathbf{I}$ the nuclear spin, $I$ the nuclear quantum number, $\omega_e$ the Larmor frequency of a bare electron, and $\varphi = \rho/4 + \mathbf{S} \cdot \rho \mathbf{S}$ the nuclear part of $\rho$, which is not directly affected by SE collisions. The SE rate is $\Gamma = n\sigma v$, where $n$ is the atomic number density, $\sigma$ the SE cross-section, and $v$ the relative speed of atoms. Without SE, the hyperfine spins, the atomic spin $\mathbf{F} \equiv \mathbf{I} + \mathbf{S}$ on the hyperfine multiplet, precess about the field in the opposite senses at the same frequency $\omega_0 = \omega_e/(2I+1)$. We will assume the atom to be $^{87}$Rb and set $I = 3/2$ hereafter.

For a linearly polarized probe beam propagating along the z-direction, its FR is $\phi = -nl\Phi$, where $l$ is the cell length, and $\Phi$ is the FR cross-section given by

$$\Phi = \chi_a F_{z,a} + \chi_b F_{z,b}. \quad (2)$$

The hyperfine spin $F_{z,F}$ represents $F_z$ for the ground state hyperfine multiplet $F$, $a = I + 1/2$, $b = I - 1/2$, and $\chi_F(v)$ is the detuning factor for multiplet $F$ given by [50]

$$\chi_a(v) = \frac{\pi r_e c f_{D1}}{2I+1}\left[\frac{1}{4}L(\Delta v_{aa'}) + \frac{3}{4}L(\Delta v_{ab'})\right],$$
$$\chi_b(v) = -\frac{\pi r_e c f_{D1}}{2I+1}\left[\frac{5}{4}L(\Delta v_{ba'}) - \frac{1}{4}L(\Delta v_{bb'})\right], \quad (3)$$

where $r_e$ is the classical radius of the electron, $c$ is the speed of light, $f_{D1} = 0.34$ the oscillator strength of the $^{87}$Rb D1 line, $\Delta v_{FF'} \equiv v - v_{FF'}$ is the detuning from the $F$-$F'$ resonance, and $L(\Delta v_{FF'})$ is the real (dispersive) part of the Voigt profile of the resonance [51]. The OSN is given by the variance of $\phi$ averaged over all the probed atoms,

$$\overline{\delta^2 \phi} = n^2 l^2 \langle \Phi^2 \rangle \frac{1}{nlA_P} = \frac{nl}{A_P}\langle \Phi^2 \rangle, \quad (4)$$

where $A_P$ is the beam area, and $nlA_P$ is the total number of atoms being probed simultaneously.

We consider the condition $W \gg \Gamma, \omega_0$, under which the total SN power around $\omega_0$ is nearly independent of $\omega_0$ [52]. This condition also allows Eq.(1) to be solved by treating the Zeeman and the SE term as perturbations to the hyperfine interaction [53,4].

For weak SE coupling (i.e., $\omega_0 \gg \Gamma_{SE}$), Eq.(1) yields two eigenobservables (EOs), $F_{z,a}$ and $F_{z,b}$ with respective independent decay rates $\gamma_a = \Gamma/8$ and $\gamma_b = 5\Gamma/8$. The variance of $\Phi$ is then

$$\langle \Phi^2 \rangle = \chi_a^2 \langle F_{z,a}^2 \rangle + \chi_b^2 \langle F_{z,b}^2 \rangle, \quad (5)$$

where the variance of $F_{z,F}$ at the thermal state is given by

$$\langle F_{z,F}^2 \rangle = \frac{2F+1}{2(2I+1)}\frac{F(F+1)}{3} = \begin{cases} 5/4, & F = 2 \\ 1/4, & F = 1 \end{cases}. \quad (6)$$

For strong SE coupling (i.e., $\omega_0 = 0$ for simplicity), the eigensolution of Eq.(1) returns two independent EOs [4,7]

$$F_{z+} = \frac{1}{6}(F_{z,a} + F_{z,b}), \quad F_{z-} = \frac{1}{6}(-F_{z,a} + 5F_{z,b}), \quad (7)$$

with respective decay rates $\gamma_+ \approx 0$ and $\gamma_- = 3\Gamma/4$. We have modified the usual eigensolution by replacing any zero-valued expectation value of an observable with the observable itself to allow it to fluctuate. The EO $F_{z+}$ represents a state of positive HSC, which is the well-known spin-temperature state. In contrast, $F_{z-}$ represents a state of negative HSC, in which the two hyperfine spins point in opposite directions [4]. Using Eq.(7), we express $F_{z,F}$ in terms of $F_{z\pm}$ in Eq.(2) and obtain $\Phi = \Phi_+ + \Phi_-$, where

$$\Phi_+ = (5\chi_a + \chi_b)F_{z+}, \quad \Phi_- = (-\chi_a + \chi_b)F_{z-}. \quad (8)$$

Using the fact that $\langle F_{z,a}F_{z,b}\rangle = \langle F_{z,a}\rangle_{F=a}\langle F_{z,b}\rangle_{F=b} = 0$ at the thermal state, we get $\langle F_{z+}F_{z-}\rangle = 0$ and $\langle \Phi_+ \Phi_-\rangle = 0$. Therefore, the total variance of $\Phi$ in a zero-field is

$$\langle \Phi^2 \rangle_{ZF} = \langle \Phi_+^2 \rangle + \langle \Phi_-^2 \rangle$$
$$\langle \Phi_+^2 \rangle = \frac{(5\chi_a + \chi_b)^2}{36}\left[\langle F_{z,a}^2\rangle + \langle F_{z,b}^2\rangle\right], \quad (9)$$
$$\langle \Phi_-^2 \rangle = \frac{(\chi_a - \chi_b)^2}{36}\left[\langle F_{z,a}^2\rangle + 25\langle F_{z,b}^2\rangle\right]$$

Such HSCs of noise in a single alkali vapor is analogous to the cross-correlation spin noise of the different alkali species in a mixed vapor [39]. While Eq.(9) and (5) seem to be very different, after plugging in the value of $\langle F_{z,F}^2\rangle$ from Eq.(6), we find

$$\langle \Phi^2(v)\rangle_{ZF} = \langle \Phi^2(v)\rangle = (5\chi_a^2 + \chi_b^2)/4. \quad (10)$$

It shows that the total OSN power is independent of the SE coupling strength. We define the following power ratios

$$\xi_\pm(v) \equiv \frac{\langle \Phi_\pm^2(v)\rangle}{\langle \Phi^2(v)\rangle}, \quad \xi \equiv \xi_+(v) + \xi_-(v). \quad (11)$$

While $\xi = 1$ according to Eq(10), $\xi_+(v)$ and $\xi_-(v)$ varies complementarily between 0 and 1 with the detuning. There are two 'polar' frequencies $v_\pm$ defined by $\chi_a(v_+) = \chi_b(v_+)$ and $5\chi_a(v_-) = -\chi_b(v_-)$, at which we find $\xi_\pm(v_\pm) = 1$ and $\xi_\mp(v_\pm) = 0$, namely, the zero-field OSN is entirely $\Phi_+$ or $\Phi_-$ polarized, respectively. For the case of far detunings

(i.e., $|\Delta\nu_{FF'}| \gg W$), $\chi_b \approx -\chi_a$, we have $\xi_+(\infty) = 4/9$ and $\xi_-(\infty) = 5/9$.

The final autocorrelation function of $\Phi$ in a zero-field is

$$\langle \Phi(t_0)\Phi(t_0+\tau)\rangle_{ZF} = \left(\langle \Phi_+^2\rangle + \langle \Phi_-^2\rangle e^{-\gamma_-\tau}\right)e^{-\gamma_W\tau}, \quad (12)$$

where the last factor is appended to approximately account for other small relaxations neglected in the master equation.

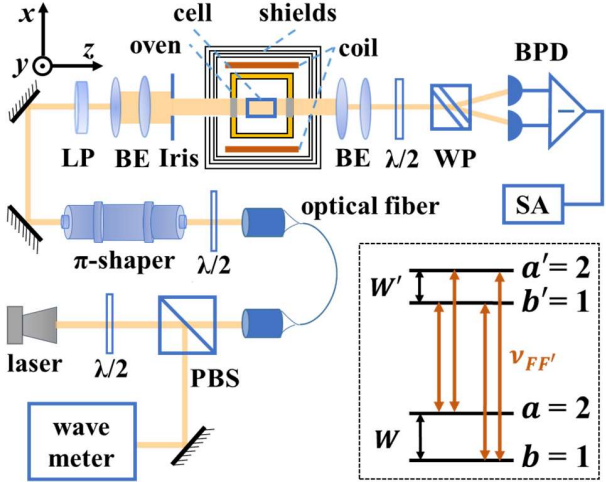

FIG. 1 (color online) Experimental setup. PBS, polarizing beam splitter; $\lambda/2$, half-wave plate; $\pi$-shaper, Gaussian to flattop beam profile convertor; LP, linear polarizer; BE, beam expander; Iris, adjustable rectangular aperture; WP, Wollaston prism; BPD, balanced-photo-detector; SA, spectrum analyzer. The insect is the hyperfine structure of the $^{87}$Rb D1 line. The ground and the excited state hyperfine splittings are $W = 6.8$ GHz and $W' = 0.8$ GHz, respectively.

Our experimental setup is shown in FIG. 1. We use a fused quartz cell with an inner dimension of 23 mm length and $8\times 8$ mm$^2$ cross-section. The cell is internally coated with an antirelaxation film of octadecyl-trichlorosilane [54,55] and filled with enriched $^{87}$Rb. It is placed in a ceramic oven wrapped and heated by a nonmagnetic wire passing an AC of 71 kHz. The cell's stem storing the Rb reservoir is kept a few degrees cooler than the body. A set of Helmholtz coil driven by a low noise current source (ADC6156) produces a dc magnetic field $\mathbf{B}_0$ in the $x$ direction. Another set of coils driven by a homemade current-pulse generator produces a pulse-modulated-field $\mathbf{B}_p$ in the same direction. The whole setup is contained in a four-layer μ-metal shield. We probe the OSN by a linear polarized diode laser (Toptica DLpro) which can be detuned between $\pm 60$ GHz about the $^{87}$Rb D1 line and is monitored by a wavemeter (Bristol 771B). We convert the beam profile to a flattop square to match the cell's cross-section by a single-mode optical fiber, a πShaper, a beam expander, and a rectangular iris. The initial plane of polarization of the probe is set at the magic angle (i.e., 54.7° to $\mathbf{B}_0$) to eliminate the tensor light shift [56,57]. Behind the cell, a Wollaston prism and a balanced photodetector measure the probe's FR. The FR signal is fed into an SRS760 spectrum analyzer (SA) to obtain its power spectrum density (PSD), which is averaged for about 5 to 30 minutes depending on the signal size. Finally, we extract the SN-induced FRN by subtracting from the former PSD the background measured at $B_0 = 0.5$ G, in which the spin resonance is shifted outside the SA's bandwidth.

To understand the experiment data, we review the property of SN spectra in a πPM-field with modulation rate $\nu_p$ and duty cycle $d$ [48]. The spectrum contains a series of equal-width resonant harmonics peaked at odd integer ($n_p$) multiples of $\nu_p/2$ with amplitude $\propto 1/n_p^2$. The total SN power is equal to the power of the first harmonic ($n_p = 1$) divided by $\pi^2/8 \approx 0.81$. Since a πPM-field is equivalent to a zero-field for isotropic spin interactions, the SE broadening of each resonance is reduced to $d$ times its value in a large dc field.

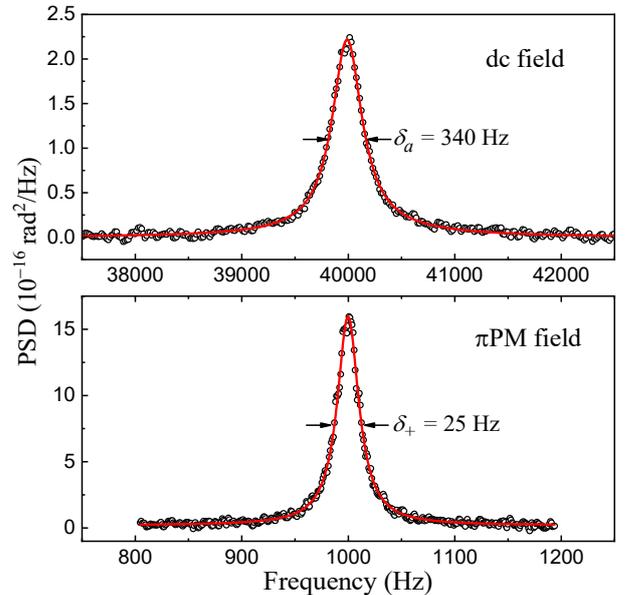

FIG. 2 (color online) The OSN spectra in a dc-field and a πPM-field. The cell's stem temperature $T = 108.2°$C; the probe's power is 0.19 mW; the probe is red detuned 14.1 GHz from the $ab'$ transition. The data averaged for 320 s (black circle) are fit by the theory (solid red line). The frequency resolution of the dc-field spectrum is ~16 Hz. The πPM-field spectrum is scanned about its first harmonic with frequency resolution ~1 Hz. ($\nu_p = 2$ kHz, $d = 0.14\%$)

In FIG. 2, we compare the noise spectra measured in a dc-field and a πPM-field corresponding to the weak and strong SE coupling regime, respectively. The dc-field spectrum can be fit by the sum of two Lorentzians representing the uncorrelated hyperfine spin noise, $\langle\Phi_a^2\rangle$ and $\langle\Phi_b^2\rangle$, whose resonant frequencies are set 159 Hz apart during the fitting to account for the nuclear spin Zeeman effect. The full width of the $\langle\Phi_a^2\rangle$ peak is $\delta_a = \delta_w + \gamma_a/\pi$, where $\delta_w$ represents the wall relaxation broadening. The total fitted area gives the value of $\langle\Phi^2\rangle$. For the πPM-field spectrum, we fit the first harmonic by a single Lorentzian. The fitted area divided by 0.81 gives the value of $\langle\Phi_+^2\rangle$. The 25 Hz fitted linewidth is equal to $\delta_w$ as the SE broadening is suppressed by a factor of $1/d \approx 700$. From $\delta_a$ and $\delta_w$, we obtain $\gamma_a/\pi \approx 315$ Hz. We rescanned the πPM-field spectrum of FIG. 2 in FIG. 3, which displays two more harmonics of the spin resonance. The enlarged base of the spectrum reveals the broad structure of $\langle\Phi_-^2\rangle$, which is fit by the sum of three Lorentzians obeying the general property of πPM-field spectra described earlier. The fitted area of the first harmonic divided by 0.81 gives the value of $\langle\Phi_-^2\rangle$. The fitted full-width is 1.6 kHz, while the theoretical width should be $\gamma_-/\pi = 6\gamma_a/\pi = 1.9$ kHz. This small difference might be caused by the blending of the narrow spectrum of $\langle\Phi_+^2\rangle$ into that of $\langle\Phi_-^2\rangle$.

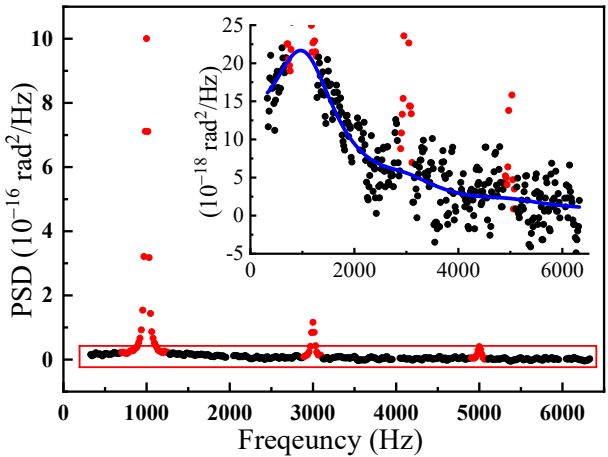

FIG. 3 (color online) A wider scan of the πPM-field OSN spectrum at the same condition as FIG. 2. The electronic interference spikes at integer multiples of $\nu_p$ are trimmed. We can see two more resonant harmonics of the pulsed spin precession. Since the SA can plot only 400 points, the frequency resolution of this scan is 15.6 Hz, close to the line width of each peak, which causes the 1st peak here to be lower than the one in FIG. 2. The inset shows an enlarged view of the base of the spectrum. The solid blue line is the fitting with the narrow peaks (red points) being masked.

From FIG. 2 and 3, we get $\xi_+ = 0.58(3)$ and $\xi = 1.01(6)$. We find no systematic dependence of $\xi_+$ on the probe power from 0.08 mW to 1.8 mW within 9% uncertainty. In FIG. 4 (a), we plot the dependence of $\xi_+$ and $\xi$ on the detuning relative to the c.m. of the D1 line. The experiment results agree with the numerical calculation of the theory very well. We find the polar frequencies to be 6.3 GHz and 0.6 GHz, at which $\langle\Phi_+^2\rangle$ and $\langle\Phi_-^2\rangle$ vanishes, respectively. The noise spectrum at 6.3 GHz detuning is shown in FIG. 4(b).

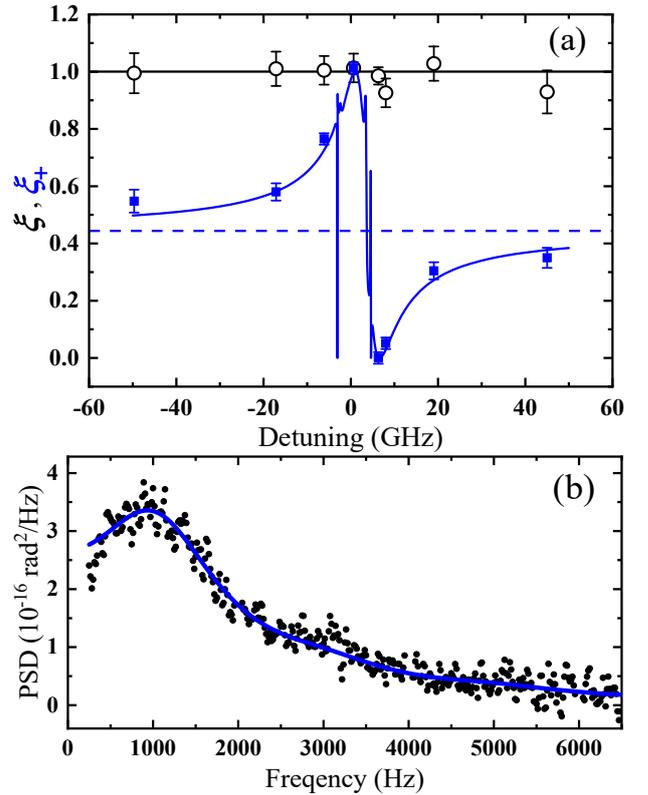

FIG. 4 (color online) (a) The detuning dependence of $\xi_+$ and $\xi$ at $T = 108.2$ °C. For $\xi_+$: experiment (blue square), theory (solid blue line). The dashed blue line is the far detuning limit $\xi_+(\infty)$. For $\xi$: experiment (black circle), theory (solid black line). (b) The OSN spectrum measured at the polar frequency $\nu_- = 6.3$ GHz where $\langle\Phi_+^2\rangle$ vanishes. The fitting (solid blue line) returns a full width of 1.9 kHz.

Finally, we show the measurement of $\xi_+(\nu)$ for different SE rates in FIG. 5. The SE rate at 125.7°C is about three times bigger than that at 107.3°C, but their $\xi_+(\nu)$ coincide very well. On the other hand, at 50.2°C, the SE rate is lower than the wall relaxation rate. Thus the strong SE coupling condition is broken, and no obvious HSCs can be observed.

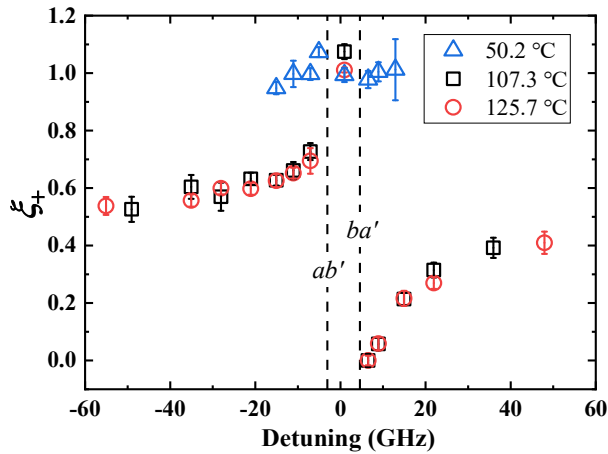

FIG. 5 (color online) The probe detuning dependence of $\xi_+$. We took three sets of data for the cell's temperature of 107.3°C (Black square), 125.7°C (red circle), and 50.2°C (blue triangle). The error bar of each point represents the spread of values of three repeated measurements. The vertical dashed lines mark frequencies of the *ab'* and the *ba'* transitions.

In summary, we study the zero-field OSN spectra of a hot alkali vapor at thermal equilibrium. In this strong SE coupling regime or the SERF regime, the spectrum consists of two components corresponding to a positive and a negative HSC. The total OSN power is independent of the SE coupling strength. The power percentages of the positive and the negative HSC components vary in a complementary way between 0 and 1 with the probe's detuning frequency. There are two polar frequencies at which the OSN spectrum is 100% polarized with one HSC component. Note, the opposite HSC component still exists in the spin system, but it is invisible to the FR probe. These polar frequencies allow the probe light to selectively interact with only one type of hyperfine spin correlation and open special windows on the light-atom quantum interface. At far detunings, which is the most popular condition used in FR detection, the spectrum amplitude of the negative HSC component is much lower, about two orders of magnitude in our case, than that of the positive one, but it contains more than half of the total noise power. Overlooking the negative HSC noise component might lead to inaccurate estimations of spin squeezing. Lastly, our approach of deriving the OSN spectrum using the eigensolution of the density matrix master equation also applies to other spin interactions and multiple species. Thus, much knowledge from the previous works [4,7,53,58] can be readily made used of for SN calculation.

## ACKNOWLEDGMENTS

We thank M. Romalis for helpful discussions. This work is supported by the National Key Research Program of China (2016YFA0302000), the National Natural Science Foundation of China (91636102), the Natural Science Foundation of Shanghai (16ZR1402700).

[1]  J. P. Wittke and R. H. Dicke, *Redetermination of the Hyperfine Splitting in the Ground State of Atomic Hydrogen*, Phys. Rev. **103**, 620 (1956).
[2]  E. M. Purcell and G. B. Field, *Influence of Collisions upon Population of Hyperfine States in Hydrogen*, Astrophys. J. **124**, 542 (1956).
[3]  F. Grossetete, *Relaxation Par Collisions d'échange de Spin*, Journal De Physique **25**, 383 (1964).
[4]  W. Happer and A. C. Tam, *Effect of Rapid Spin Exchange on the Magnetic-Resonance Spectrum of Alkali Vapours*, Phys. Rev. A **16**, 1877 (1977).
[5]  M. A. Bouchiat, T. R. Carver, and C. M. Varnum, *Nuclear Polarization in He3 Gas Induced by Optical Pumping and Dipolar Exchange*, Physical Review Letters **5**, 373 (1960).
[6]  T. G. Walker and W. Happer, *Spin-Exchange Optical Pumping of Noble-Gas Nuclei*, Reviews of Modern Physics **69**, 629 (1997).
[7]  S. Appelt, A. Ben-Amar Baranga, C. J. Erickson, M. Romalis, A. R. Young, and W. Happer, *Theory of Spin-Exchange Optical Pumping of 3He and 129Xe*, Physical Review A **58**, 1412 (1998).
[8]  T. R. Gentile, P. J. Nacher, B. Saam, and T. G. Walker, *Optically Polarized He-3*, Rev. Mod. Phys. **89**, 045004 (2017).
[9]  J. Vanier and C. Audoin, *The Quantum Physics of Atomic Frequency Standards*, Vol. 1 (Institute of Physics Publishing, 1989).
[10] D. Budker and M. Romalis, *Optical Magnetometry*, Nat Phys **3**, 227 (2007).
[11] M. S. Albert, G. D. Cates, B. Driehuys, W. Happer, B. Saam, C. S. Springer Jr., and A. Wishnia, *Biological Magnetic Resonance Imaging Using Laser-Polarized 129Xe*, Nature **370**, 199 (1994).
[12] K. Ruppert, *Biomedical Imaging with Hyperpolarized Noble Gases*, Rep. Prog. Phys. **77**, 116701 (2014).
[13] B. C. Grover, *Noble-Gas NMR Detection through Noble-Gas-Rubidium Hyperfine Contact Interaction*, Phys. Rev. Lett. **40**, 391 (1978).
[14] M. A. Rosenberry and T. E. Chupp, *Atomic Electric Dipole Moment Measurement Using Spin Exchange*

*Pumped Masers of Xe-129 and He-3*, Phys. Rev. Lett. **86**, 22 (2001).

[15] W. Happer and H. Tang, *Spin-Exchange Shift and Narrowing of Magnetic Resonance Lines in Optically Pumped Alkali Vapours*, Phys. Rev. Lett. **31**, 273 (1973).

[16] J. C. Allred, R. N. Lyman, T. W. Kornack, and M. V. Romalis, *High-Sensitivity Atomic Magnetometer Unaffected by Spin-Exchange Relaxation*, Physical Review Letters **89**, 130801 (2002).

[17] I. K. Kominis, T. W. Kornack, J. C. Allred, and M. V. Romalis, *A Subfemtotesla Multichannel Atomic Magnetometer*, Nature **422**, 596 (2003).

[18] T. W. Kornack and M. V. Romalis, *Dynamics of Two Overlapping Spin Ensembles Interacting by Spin Exchange*, Physical Review Letters **89**, 253002 (2002).

[19] T. W. Kornack, R. K. Ghosh, and M. V. Romalis, *Nuclear Spin Gyroscope Based on an Atomic Comagnetometer*, Physical Review Letters **95**, 230801 (2005).

[20] G. Vasilakis, J. M. Brown, T. W. Kornack, and M. V. Romalis, *Limits on New Long Range Nuclear Spin-Dependent Forces Set with a K-He-3 Comagnetometer*, Phys. Rev. Lett. **103**, (2009).

[21] M. Smiciklas, J. M. Brown, L. W. Cheuk, S. J. Smullin, and M. V. Romalis, *New Test of Local Lorentz Invariance Using a (21)Ne-Rb-K Comagnetometer*, Phys. Rev. Lett. **107**, 171604 (2011).

[22] J. Lee, A. Almasi, and M. Romalis, *Improved Limits on Spin-Mass Interactions*, Phys. Rev. Lett. **120**, 161801 (2018).

[23] T. L. Ho, *Spinor Bose Condensates in Optical Traps*, Phys. Rev. Lett. **81**, 742 (1998).

[24] E. M. Bookjans, C. D. Hamley, and M. S. Chapman, *Strong Quantum Spin Correlations Observed in Atomic Spin Mixing*, Phys. Rev. Lett. **107**, 210406 (2011).

[25] X.-Y. Luo, Y.-Q. Zou, L.-N. Wu, Q. Liu, M.-F. Han, M. K. Tey, and L. You, *Deterministic Entanglement Generation from Driving through Quantum Phase Transitions*, Science **355**, 620 (2017).

[26] Y.-Q. Zou, L.-N. Wu, Q. Liu, X.-Y. Luo, S.-F. Guo, J.-H. Cao, M. K. Tey, and L. You, *Beating the Classical Precision Limit with Spin-1 Dicke States of More than 10,000 Atoms*, Proc. Natl. Acad. Sci. U. S. A. **115**, 6381 (2018).

[27] L. Pezze, A. Smerzi, M. K. Oberthaler, R. Schmied, and P. Treutlein, *Quantum Metrology with Nonclassical States of Atomic Ensembles*, Rev. Mod. Phys. **90**, 035005 (2018).

[28] K. Jensen, W. Wasilewski, H. Krauter, T. Fernholz, B. M. Nielsen, J. M. Petersen, J. J. Renema, M. V. Balabas, M. Owari, M. B. Plenio, A. Serafini, M. M. Wolf, C. A. Muschik, J. I. Cirac, J. H. Müller, and E. S. Polzik, *Quantum Memory, Entanglement and Sensing with Room Temperature Atoms*, Journal of Physics: Conference Series **264**, 012022 (2011).

[29] G. Vasilakis, H. Shen, K. Jensen, M. Balabas, D. Salart, B. Chen, and E. S. Polzik, *Generation of a Squeezed State of an Oscillator by Stroboscopic Back-Action-Evading Measurement*, Nat. Phys. **11**, 389 (2015).

[30] H. Bao, J. Duan, S. Jin, X. Lu, P. Li, W. Qu, M. Wang, I. Novikova, E. E. Mikhailov, K.-F. Zhao, K. Mølmer, H. Shen, and Y. Xiao, *Spin Squeezing of 10^11 Atoms by Prediction and Retrodiction Measurements*, Nature **581**, 7807 (2020).

[31] I. K. Kominis, *Sub-Shot-Noise Magnetometry with a Correlated Spin-Relaxation Dominated Alkali-Metal Vapor*, Phys. Rev. Lett. **100**, 093602 (2008).

[32] J. Kong, R. Jiménez-Martínez, C. Troullinou, V. G. Lucivero, G. Tóth, and M. W. Mitchell, *Measurement-Induced, Spatially-Extended Entanglement in a Hot, Strongly-Interacting Atomic System*, Nature Communications **11**, 2415 (2020).

[33] K. Mouloudakis and I. K. Kominis, *Spin-Exchange Collisions in Hot Vapors Creating and Sustaining Bipartite Entanglement*, Phys. Rev. A **103**, L010401 (2021).

[34] A. Dantan, G. Reinaudi, A. Sinatra, F. Laloë, E. Giacobino, and M. Pinard, *Long-Lived Quantum Memory with Nuclear Atomic Spins*, Phys. Rev. Lett. **95**, 123002 (2005).

[35] O. Katz, R. Shaham, E. S. Polzik, and O. Firstenberg, *Long-Lived Entanglement Generation of Nuclear Spins Using Coherent Light*, Phys. Rev. Lett. **124**, 043602 (2020).

[36] O. Katz, R. Shaham, and O. Firstenberg, *Quantum Interface for Noble-Gas Spins Based on Spin-Exchange Collisions*, ArXiv:1905.12532 [Physics, Physics:Quant-Ph] (2021).

[37] R. Shaham, O. Katz, and O. Firstenberg, *Strong Coupling of Alkali Spins to Noble-Gas Spins with Hour-Long Coherence Time*, ArXiv:2102.02797 [Quant-Ph] (2021).

[38] A. T. Dellis, M. Loulakis, and I. K. Kominis, *Spin-Noise Correlations and Spin-Noise Exchange Driven by Low-Field Spin-Exchange Collisions*, Phys. Rev. A **90**, 032705 (2014).

[39] D. Roy, L. Y. Yang, S. A. Crooker, and N. A. Sinitsyn, *Cross-Correlation Spin Noise Spectroscopy of Heterogeneous Interacting Spin Systems*, Sci Rep **5**, 9573 (2015).

[40] K. Mouloudakis, M. Loulakis, and I. K. Kominis, *Quantum Trajectories in Spin-Exchange Collisions Reveal the Nature of Spin-Noise Correlations in Multispecies Alkali-Metal Vapors*, Phys. Rev. Research **1**, 033017 (2019).

[41] E. B. Aleksandrov and V. S. Zapassky, *Magnetic-resonance in the Faraday-rotation noise spectrum*, Zhurnal Eksperimentalnoi Teor. Fiz. **81**, 132 (1981).

[42] S. A. Crooker, D. G. Rickel, A. V. Balatsky, and D. L. Smith, *Spectroscopy of Spontaneous Spin Noise as a Probe of Spin Dynamics and Magnetic Resonance*, Nature **431**, 49 (2004).

[43] V. S. Zapasskii, *Spin-Noise Spectroscopy: From Proof of Principle to Applications*, Adv. Opt. Photonics **5**, 131 (2013).

[44] N. A. Sinitsyn and Y. V. Pershin, *The Theory of Spin Noise Spectroscopy: A Review*, Rep. Prog. Phys. **79**, 106501 (2016).

[45] J. L. Sorensen, J. Hald, and E. S. Polzik, *Quantum Noise of an Atomic Spin Polarization Measurement*, Phys. Rev. Lett. **80**, 3487 (1998).

[46] V. S. Zapasskii, A. Greilich, S. A. Crooker, Y. Li, G. G. Kozlov, D. R. Yakovlev, D. Reuter, A. D. Wieck, and M. Bayer, *Optical Spectroscopy of Spin Noise*, Phys. Rev. Lett. **110**, 176601 (2013).

[47] M. E. Limes, D. Sheng, and M. V. Romalis, *3He-129Xe Comagnetometery Using 87Rb Detection and Decoupling*, Phys. Rev. Lett. **120**, 033401 (2018).

[48] G. Zhang, Y. Wen, J. Qiu, and K. Zhao, *Spin-Noise Spectrum in a Pulse-Modulated Field*, Opt. Express **28**, 15925 (2020).

[49] Y. Tang, Y. Wen, L. Cai, and K. Zhao, *Spin-Noise Spectrum of Hot Vapor Atoms in an Anti-Relaxation-Coated Cell*, Phys. Rev. A **101**, 013821 (2020).

[50] W. Happer and B. S. Mathur, *Effective Operator Formalism in Optical Pumping*, Physical Review **163**, 12 (1967).

[51] S. J. Seltzer, Developments in Alkali-Metal Atomic Magnetometry, Princeton University, 2008.

[52] B. Mihaila, S. A. Crooker, D. G. Rickel, K. B. Blagoev, P. B. Littlewood, and D. L. Smith, *Quantitative Study of Spin Noise Spectroscopy in a Classical Gas of K-41 Atoms*, Phys. Rev. A **74**, 043819 (2006).

[53] W. Happer, *Optical Pumping*, Reviews of Modern Physics **44**, 169 (1972).

[54] K. F. Zhao, E. Ulanski, M. Schaden, and Z. Wu, *Dwell Time Measurement of Wall Collisions of Spin Polarized Rb Atoms on Antirelaxation Coatings*, Journal of Physics: Conference Series **388**, 012046 (2012).

[55] K. J. Liao, M. L. Wang, G. Y. Zhang, and K. F. Zhao, *Time-Resolved Measurements of the Adsorption/Desorption of Rb Atoms on Octadecyltrichlorosilane Coated Surfaces*, Chin. Phys. Lett. **32**, 76801 (2015).

[56] B. S. Mathur, H. Tang, and W. Happer, *Light Shifts in the Alkali Atoms*, Physical Review **171**, 11 (1968).

[57] W. Chalupczak and R. M. Godun, *Near-Resonance Spin-Noise Spectroscopy*, Physical Review A **83**, 032512 (2011).

[58] O. Katz, O. Peleg, and O. Firstenberg, *Coherent Coupling of Alkali Atoms by Random Collisions*, Physical Review Letters **115**, 113003 (2015).